\documentclass[letterpaper,twocolumn,superscriptaddress,pre,aps,showpacs,
floatfix]{revtex4-1}
\usepackage[latin1]{inputenc}
\usepackage{bm}
\usepackage[usenames]{color}
\usepackage{multirow}
\usepackage{amssymb}
\usepackage{amsbsy}
\usepackage{amsmath}
\usepackage{stmaryrd}
\usepackage{graphicx}
\usepackage{epsfig}
\usepackage{placeins}
\usepackage{ulem}
\usepackage{bbold}
\usepackage{bbm}
\usepackage{braket}
\usepackage[colorlinks,linkcolor=blue,citecolor=blue,urlcolor=blue]{hyperref}
\usepackage{blindtext}

\usepackage{array}

\makeatletter

\begin{document}

\title{Eigenstate thermalization hypothesis beyond standard 
indicators: Emergence of random-matrix behavior at small frequencies}

\author{Jonas Richter}
\email{jonasrichter@uos.de}
\affiliation{Department of Physics, University of Osnabr\"uck, D-49069 
Osnabr\"uck, Germany}

\author{Anatoly Dymarsky}
\email{a.dymarsky@uky.edu}
\affiliation{Moscow Institute of Physics and Technology, 9 Institutskiy 
pereulok, Dolgoprudny, 141700, Russia}
\affiliation{Skolkovo Institute of Science and Technology, 
Skolkovo Innovation Center, Moscow, Russia}
\affiliation{Department of Physics, University of Kentucky, 
Lexington KY, United States}

\author{Robin Steinigeweg}
\affiliation{Department of Physics, University of Osnabr\"uck, D-49069 
Osnabr\"uck, Germany}

\author{Jochen Gemmer}
\email{jgemmer@uos.de}
\affiliation{Department of Physics, University of Osnabr\"uck, D-49069 
Osnabr\"uck, Germany}

\date{\today}


\begin{abstract}
Using numerical exact diagonalization, we study matrix elements of a local spin 
operator in the eigenbasis of two different 
nonintegrable quantum spin chains. Our emphasis is on the question to what 
extent local operators can be represented as random matrices and, in 
particular, to what extent matrix elements can be considered as uncorrelated. 
As a main result, we show that the eigenvalue distribution of 
band submatrices at a fixed energy density is a sensitive probe of the 
correlations between matrix elements. We find that, on the scales where the 
matrix elements are in a good agreement with all standard indicators of the 
eigenstate thermalization hypothesis, the eigenvalue distribution 
still exhibits clear signatures of the original operator, implying correlations 
between matrix elements.
Moreover, we demonstrate that at much smaller energy scales, the eigenvalue 
distribution approximately assumes the universal semicircle shape, indicating 
transition to the random-matrix behavior, and in particular that  matrix 
elements become uncorrelated.

\end{abstract}

\maketitle


\section{Introduction}

Questions of equilibration and thermalization in isolated quantum 
many-body systems have experienced an upsurge of interest both from 
the theoretical and the experimental side over the last decades 
\cite{polkovnikov2011, gogolin2016, dalessio2016}. 
In this context, the eigenstate thermalization 
hypothesis (ETH) has been established as a key concept to explain the 
emergence of thermodynamic behavior, by assuming a certain matrix structure of 
physical operators ${\cal O}$ in the eigenbasis of generic Hamiltonians 
${\cal H}$ \cite{deutsch1991, srednicki1994, rigol2005}. 
Specifically, let ${\cal O}_{mn} = \bra{m}{\cal O}\ket{n}$ denote the matrix 
element of ${\cal O}$ within the eigenstates $\ket{m}$ and 
$\ket{n}$ of ${\cal H}$, then the ETH ansatz reads 
\cite{dalessio2016,srednicki1999}
\begin{equation}\label{Eq::ETH}
 {\cal O}_{mn} = 
O(\bar{E})\delta_{mn}+\Omega^{-\tfrac{1}{2}}(\bar{E})f(\bar{E},
\omega)r_ {mn}\ , 
\end{equation}
where $\omega= E_m - E_n$ 
is the difference between the eigenenergies $E_m$ and $E_n$ with mean 
energy $\bar{E} = 
(E_m + E_n)/2$, $O(\bar{E})$ and $f(\bar{E}, \omega)$ are smooth 
functions of their arguments, and $\Omega(\bar{E})$ is the 
density of states. Furthermore, the $r_{mn} = r_{nm}^\ast$ are 
conventionally assumed to be \mbox{(pseudo-)}random  Gaussian 
variables with zero mean and unit variance. (For earlier works, 
see also \cite{Feingold1986, Feingold1991}.)  
While the ETH is an assumption and a formal proof is 
absent, its validity (including the Gaussian distribution of the $r_{nm}$) has 
been numerically confirmed for a variety of models and 
observables 
\cite{santos2010, beugeling2014, beugeling2015, kim2014, steinigeweg2013, 
Mondaini2016, mondaini2017,  
jansen2019, LeBlond2019}. Generally, the ETH is believed to hold for 
nonintegrable 
models and physical (for instance, spatially 
local) observables. 
In contrast, the ETH is violated in integrable systems due to their extensive 
number of integrals of motion \cite{essler2016}, as well as in strongly 
disordered models 
which undergo a transition to a many-body localized phase in one dimension
\cite{nandkishore2015}. In these cases, the 
off-diagonal matrix elements $r_{nm}$ deviate from the Gaussian 
distribution \cite{LeBlond2019, Luitz2016}. In addition, 
models exhibiting a weaker violation of the ETH, such as, e.g., models 
featuring so-called 
``quantum scars'', where rare less entangled states are embedded in an 
otherwise 
thermal spectrum, have recently attracted a 
significant amount of interest (see, e.g., \cite{Shiraishi2017, Turner2018}).

While the formulation of the ETH in Eq.\ \eqref{Eq::ETH} is 
conventional~\cite{dalessio2016}, it is to some degree incomplete with 
regard to the statistical properties of the ${\mathcal O}_{mn}$. Specifically, 
for a given ${\cal H}$ and ${\cal O}$, the matrix  elements ${\cal O}_{mn}$ are 
predetermined, and therefore the notion of \mbox{(pseudo-)}randomness of the 
$r_{nm}$
needs to be carefully defined.
In the spirit of the Bohigas-Giannoni-Schmit conjecture 
\cite{Bohigas1984}, we here advocate the strongest point of 
view, that below a certain energy scale all statistical 
properties of the 
off-diagonal matrix elements would match those of a Gaussian random ensemble.
The central question of this paper is therefore to what 
extent the matrix elements ${\cal O}_{mn}$ can be represented as {\it 
independently} drawn 
random numbers? 

Clearly, all ${\cal O}_{mn}$ cannot be random in the 
strict sense as they 
are constrained by the fact that the observables  have to obey various  algebraic relations (e.g.\ 
${\mathcal O}^2= \mathbb{1}$ in case of ${\cal O}$ being a Pauli matrix acting 
on an individual spin). Furthermore, correlations between the $r_{mn}$ 
are necessary to reproduce the growth of certain four-point 
correlation functions in chaotic systems  \cite{Foini2019, Chan2019, Murthy2019}.  
Likewise, the consistency of relaxation dynamics in local systems also requires 
the $r_{mn}$ to be correlated \cite{Dymarsky2018}.  
We therefore arrive at the important conclusion that the onset 
of random-matrix behavior has to be limited 
to matrix elements ${\mathcal O}_{mn}$ within a certain energy window specified 
by the relevant energy scale $\Delta E_{\rm RMT}$. 

In this paper, we test the ETH in the case of a local spin operator in the 
eigenbasis of the 
paradigmatic spin-$1/2$ XXZ chain, 
for which we break integrability by means of (i) an additional next-nearest 
neighbor interaction or (ii) a single-site magnetic field in the center of the 
chain. (See Refs.\ \cite{Brenes2020_1, Brenes2020_2, Pandey2020, Santos2020} 
for related studies of the ETH and the emergence of quantum 
chaos in these models.) 
Going beyond the ``standard'' indicators of the ETH, we particularly 
investigate the 
existence of the 
scale $\Delta E_{\rm RMT}$ below which random matrix theory (RMT) prevails. 
To this end, we establish the eigenvalue spectrum of ${\cal O}$ as a sensitive 
probe of the correlations between the ${\cal O}_{mn}$. 
While the spectrum of the full spin operator includes only two eigenvalues 
$\pm1/2$, we particularly focus on the spectrum of band submatrices at a 
fixed energy 
density $\bar E$ where the ${\cal O}_{mn}$ are restricted
to a narrow band $|E_n-E_m|\leq \omega_c$.
For such band submatrices, we demonstrate that the ${\cal O}_{mn}$ are in 
convincing agreement 
with conventional indicators of the ETH in the following sense:
(i) the diagonal matrix 
elements form a ``smooth'' function of energy $O({\bar E})$, (ii) the 
off-diagonal matrix 
elements follow a Gaussian distribution with a variance $f^2(\bar{E},\omega)$ 
that depends smoothly 
on the mean energy and 
respective energy difference, 
and 
(iii) the ratio between the variances of diagonal and off-diagonal elements for 
small 
$\omega$ takes on the value predicted by RMT. 
However, despite (i) - (iii) being satisfied, we find that the eigenvalues 
of the band
submatrices for $\omega_c$ larger than a certain value $\Delta E_{\rm 
RMT}$ still exhibit clear signatures of the original operator, implying 
correlations between matrix elements. 
At the same time, when the bandwidth is sufficiently decreased, the spectrum 
takes on an approximately 
semicircular shape, marking the transition where genuine random-matrix behavior 
occurs. 

\subsection{Outline and reader's guide}

While our main goal is to demonstrate the 
existence of the scale $\Delta E_\text{RMT}$, this paper includes a 
detailed discussion of standard indicators of the ETH extensively studied in 
this context. A reader already familiar with numerical 
studies of the ETH, including works \cite{dalessio2016, deutsch1991, 
srednicki1994, rigol2005, srednicki1999, santos2010, beugeling2014, 
beugeling2015, kim2014, steinigeweg2013, 
Mondaini2016, mondaini2017,  
jansen2019, LeBlond2019}, may go directly to the relevant sections concerned 
with the 
investigation of $\Delta E_\text{RMT}$. This paper is 
structured as follows. In Sec.\ \ref{Sec::Setup}, we introduce 
the models and observables and describe our approach to 
study the ETH. In particular, the spectrum of band 
submatrices as a probe for the onset of random-matrix behavior is discussed 
in Sec.\ \ref{Sec::IndiRMT}. Our numerical results are then presented in Sec.\ 
\ref{Sec::Results}. Specifically, we present data for ``standard'' indicators 
of the ETH in Sec.\ \ref{Sec::Standard}, while the main results 
concerning the existence of the scale $\Delta 
E_\text{RMT}$ are analyzed in Sec.\ \ref{Sec::EVs}. A summary and 
discussion 
is given in Sec.\ \ref{Sec::Summary}, where we put our findings 
into context with previous studies of the ETH and outline future directions of 
research.

\section{Setup}\label{Sec::Setup}

\subsection{Models and observable}
\begin{figure}[tb]
\centering
\includegraphics[width=\columnwidth]{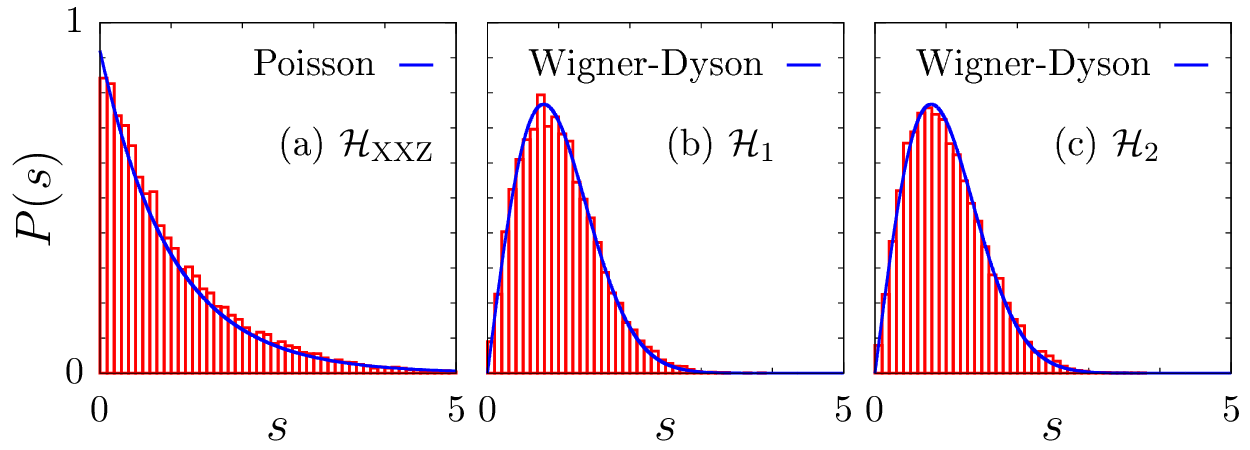}
\caption{(Color online) Level-spacing distribution $P(s)$ of (a) ${\cal 
H}_\text{XXZ}$, (b) ${\cal H}_1$, and (c) ${\cal H}_2$. The parameters are 
chosen as $\Delta = 1.5$, $\Delta^\prime = 1.2$, $h_{L/2} = 1$, and $L = 18$. 
Correct extraction of $P(s)$ requires unfolding of the 
spectrum.}
 \label{Fig1}
\end{figure}

In order to test the ETH ansatz \eqref{Eq::ETH}, we consider different 
(integrable and nonintegrable) quantum spin chains. A convenient starting point 
is provided by the  
one-dimensional XXZ model with open boundary conditions, 
\begin{equation}\label{Eq::XXZ}
 {\cal H}_\text{XXZ} = \sum_{\ell = 1}^{L-1} S_\ell^x S_{\ell+1}^x + S_{\ell}^y 
S_{\ell+1}^y + \Delta S_{\ell}^z S_{\ell+1}^z\ ,   
\end{equation}
where $L$ denotes the number of lattice sites, 
$S_\ell^{x,y,z}$ are spin-$1/2$ operators at site $\ell$, and $\Delta$ is an 
anisotropy in the $z$ direction. (In the following, we set the anisotropy to 
$\Delta = 1.5$.) While ${\cal H}_\text{XXZ}$ is integrable in 
terms of the Bethe ansatz, we break integrability by  
either an additional next-nearest neighbor interaction of strength~$\Delta'$ 
\cite{steinigeweg2013, Rigol2010, richter2018_3},
\begin{equation}\label{Eq::H1}
 {\cal H}_{1} = {\cal H}_\text{XXZ} + \Delta^\prime \sum_{\ell = 1}^{L-2} 
S_\ell^z S_{\ell+2}^z\ , 
\end{equation}
or by means of a single-site magnetic field $h_{L/2}$ in the center of the 
chain \cite{Barisic2009, Brenes2020_1, Brenes2020_2, Pandey2020, Santos2020, 
Santos2004},
\begin{equation}\label{Eq::H2}
  {\cal H}_{2} = {\cal H}_\text{XXZ} + h_{L/2} S_{L/2}^z\ . 
\end{equation}
Note that, although not written explicitly in Eqs.\ 
\eqref{Eq::XXZ}-\eqref{Eq::H2}, we furthermore always include a small magnetic 
field at the 
first 
lattice site, $h_1S_1^z$ with $h_1 = 0.1$, which breaks the spin-flip and 
reflection symmetry of the model. 

While ${\cal H}_\text{XXZ}$ and 
${\cal H}_{1,2}$ conserve the total magnetization $S^z = \sum_\ell S_\ell^z$, 
all results presented in this paper are obtained for the largest symmetry 
subspace which corresponds to $S^z = 0$ and has dimension 
\begin{equation}
 {\cal D} = \binom{L}{L/2}= \frac{L!}{(L/2)!(L/2)!}\ . 
\end{equation}
For $L = 18$, which is the largest 
system size we 
can treat numerically, we have ${\cal D} = 48620$. 
Moreover, our simulations are performed for a representative choice 
of the integrability-breaking parameters, i.e., $\Delta^\prime = 1.2$ and 
$h_{L/2} = 1$, for which both ${\cal H}_1$ and ${\cal H}_2$ are robustly 
nonintegrable (see also Refs.\ \cite{steinigeweg2013, Santos2004, 
Barisic2009, Rigol2010, richter2018_3} for other 
parameter choices).

The transition from the integrable XXZ chain to the nonintegrable models ${\cal 
H}_1$ and ${\cal H}_2$ can for instance be seen from the level-spacing 
distribution 
$P(s)$ which is shown in Fig.\ \ref{Fig1}. While the level spacings follows
the
Poisson distribution in the integrable case, $P(s)$ matches the
Wigner-Dyson distribution for ${\cal H}_{1,2}$. In this 
context, let us note that the field $h_1$ at the 
edge of the chain does not break integrability of 
${\cal H}_\text{XXZ}$ \cite{Santos2004}, while the single impurity 
$h_{L/2}$ in the center of the chain induces the onset of chaos 
\cite{Barisic2009, Santos2004}.
\begin{figure}[tb]
 \centering
 \includegraphics[width=\columnwidth]{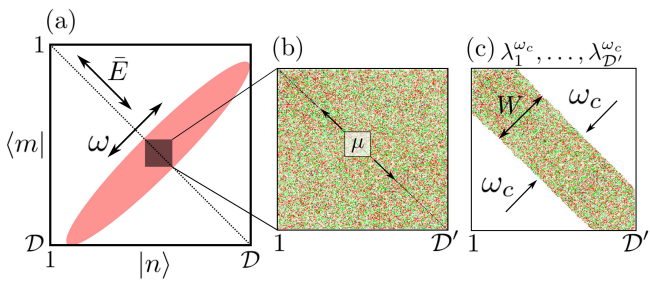}
 \caption{(Color online) (a) The ETH ansatz \eqref{Eq::ETH} is studied for the 
spin-$1/2$ operator $S_{L/2}^z$ written in 
the eigenbasis of the respective Hamiltonian ${\cal 
H}$. The oval shaded area indicates matrix elements around 
a fixed value of ${\bar E}$ where the density of states is approximately 
constant, while the smaller square-shaped shaded area 
indicates a 
submatrix in this energy window. (b) For 
square-shaped submatrices with dimension ${\cal D}^\prime < 
{\cal D}$, the 
ratio $\Sigma^2(n,\mu)$ defined in Eq.\ \eqref{Eq::SigRat} between the 
variances of diagonal and 
off-diagonal 
matrix elements is obtained for regions of size $\mu$ shifted along the 
diagonal. Note that the matrix shown here comprises actual data for the example 
of ${\cal H}_1$ and $L = 12$. (c) We introduce a cutoff frequency $\omega_c$, 
where off-diagonal 
matrix elements are set to zero,
${\cal O}_{mn} = 0$, if $|E_m - E_n| > \omega_c$, resulting in a band matrix 
with relative bandwidth $W/{\cal D}^\prime$. We study how the 
distribution of eigenvalues 
$\lambda_1^{\omega_c},\dots,\lambda_{{\cal D}'}^{\omega_c}$ of the 
submatrix evolves upon reducing $\omega_c$.}
 \label{Fig2}
\end{figure}

For nonintegrable models such as ${\cal H}_{1,2}$, it is a general 
expectation that the matrix elements of physical observables ${\cal O}$
will follow the eigenstate 
thermalization hypothesis \cite{dalessio2016}.
In this paper, we test the ETH for the case of a local spin-$1/2$ operator 
acting on 
the 
central lattice site of the chain,
\begin{equation}\label{Eq::Obs}
  {\cal O} = S_{L/2}^z\ . 
\end{equation}
Specifically, we employ full exact diagonalization to obtain the matrix 
elements ${\cal 
O}_{mn}$.  
Note that the indices $m$ and $n$ always 
refer to the eigenbasis of ${\cal H}$. The 
${\cal O}_{mn}$ are real numbers for the chosen operator 
and the symmetry subspace.

\subsection{Testing the ETH and the onset of 
RMT}\label{Sec::TestETH}

In the following, we 
introduce the quantities studied in this paper. An accompanying 
sketch is provided in Fig.~\ref{Fig2}. A reader 
familiar with the ``standard'' indicators of the ETH may directly 
go to Sec.\ \ref{Sec::IndiRMT}. 

\subsubsection{Indicators of diagonal ETH}

The ETH ansatz \eqref{Eq::ETH} consists of the
diagonal part and the off-diagonal part. The diagonal part of the ETH asserts 
that the function $O(\bar{E})$ becomes ``smooth'' in 
the thermodynamic limit $L \to \infty$. In particular, the eigenstate-to-eigenstate 
fluctuations ${\cal O}_{mm} - {\cal O}_{m+1m+1}$ should rapidly decrease 
with the system size $L$.  
One way to test this statement is to study the variance 
$\sigma^2_\text{d}(\bar{E})$ of the diagonal matrix elements,  
\begin{equation}\label{Eq::VarDiag}
\sigma^2_\text{d}(\bar{E}) = \frac{1}{N_{\bar{E}}} 
\sum_{m} [{\cal O}_{mm}]^2 - 
\left(\frac{1}{N_{\bar{E}}}\sum_{m} {\cal O}_{mm}\right)^2\ , 
\end{equation}
where the sum runs over all $N_{\bar{E}}$ eigenstates $\ket{m}$ with 
eigenenergies $E_m \in [\bar{E}-\Delta E/2,\bar{E}+\Delta E/2]$ in a 
microcanonical
energy window around a fixed $\bar{E}$. For nonintegrable systems including 
our cases, it has been found that $\sigma^2_\text{d}(\bar{E})$ decreases 
exponentially with increasing $L$, while the scaling for integrable models is 
polynomial see, e.g.,  \cite{dalessio2016, beugeling2014, steinigeweg2013, 
alba2015}.  

\subsubsection{Indicators of off-diagonal ETH}

Next, in order to test the off-diagonal part of the ETH, we consider 
matrix elements 
${\cal O}_{mn}$ in a (sufficiently narrow) energy window around a fixed 
$\bar{E}$, where $\Omega(\bar{E})$ is approximately constant, which facilitates 
the analysis of the $\omega$ dependence of $f(\bar{E},\omega)$ 
and of the distribution of the ${\cal O}_{mn}$, see Fig.\ \ref{Fig2}~(a). 
A useful quantity in this 
context is the average over matrix elements in a small $\omega$ 
interval, which we denote in this paper by an overline. For instance, the 
average over $|{\cal O}_{mn}|^2$ in an interval of width  
$\Delta \omega \ll \omega$ (with 
fixed $\bar{E}$) is given by
\begin{equation}
 \overline{|{\cal O}_{mn}|^2}(\omega) = 
\frac{1}{N_\omega}\sum_{\substack{n,m \\ E_m-E_n \approx \omega}} 
|{\cal O}_{mn}|^2\ ,  
\end{equation}
where the sum runs over all $N_\omega$  matrix elements with 
$E_m - E_n \in [\omega -\Delta\omega/2,\omega+\Delta \omega/2]$. Plotting 
$\overline{|{\cal O}_{mn}|^2}(\omega)$ versus $\omega$ yields the function 
$f^2(\bar{E},\omega)$, cf.\ Eq.\ 
\eqref{Eq::ETH}, except for an 
overall prefactor \cite{LeBlond2019, Serbyn2017, Richter2019}. 

Assuming the ${\cal O}_{mn}$ have zero mean, i.e., 
$\overline{{\cal O}_{mn}} = 0$, (which holds with a  very high accuracy), we study 
the following quantity recently introduced in Ref.\ \cite{LeBlond2019}, which is sensitive to the distribution of $r_{nm}$,
\begin{equation}\label{Eq::Gamma}
 \Gamma(\omega) = \overline{|{\cal O}_{mn}|^2}/\overline{|{\cal O}_{mn}|}^2\ . 
\end{equation}
When $\overline{{\cal O}_{mn}} = 0$, the nominator in Eq.\ 
\eqref{Eq::Gamma} coincides with the variance of the ${\cal O}_{mn}$ while the 
denominator is the squared mean of the folded distribution. 
In particular, if ${\cal O}_{mn}$ were to follow the  Gaussian distribution, 
$\Gamma(\omega) = \pi/2$. The value we find numerically in Sec.\ 
\ref{Sec::Results} is very close.
To further confirm that the distribution 
$P({\cal O}_{mn})$ of 
the ${\cal O}_{mn}$ is indeed
Gaussian, we plot the histogram of ${\cal O}_{mn}$ from narrow windows with 
fixed $\bar{E}$ and $\omega$, see Fig.~\ref{Fig5} below.

Next, we consider square-shaped submatrices of ${\cal O}$ of 
dimension 
${\cal D}^\prime < {\cal D}$ around a fixed mean energy 
$\bar{E}$. 
In 
Fig.\ \ref{Fig2}~(b), an example for such a submatrix comprising actual 
numerical data is shown. As a further check that the ${\cal 
O}_{mn}$ are normally distributed,
we calculate the ratio 
$\Sigma^2(n,\mu)$ between the variances of diagonal and off-diagonal matrix 
elements for eigenstates in (small) regions $[n-\mu/2,n+\mu/2]$ of width $\mu$ 
[cf.\ Fig.\ 
\ref{Fig2}~(b)], 
\begin{equation}\label{Eq::SigRat}
 \Sigma^2(n,\mu) = \frac{\sigma_\text{d}^2(n,\mu)}{\sigma_\text{od}^2(n,\mu)}\ .
\end{equation}
Here $\sigma_\text{d}^2(n,\mu)$ and $\sigma_\text{od}^2(n,\mu)$ are defined 
analogously to the variance in Eq.\ 
\eqref{Eq::VarDiag}, see also \cite{Note} for details. 
 
For an actual random matrix drawn from the Gaussian orthogonal ensemble 
(GOE), $\Sigma^2_\text{GOE} = 2$.  Agreement with the GOE was anticipated  in 
\cite{dalessio2016} and then verified numerically in, e.g., \cite{mondaini2017, 
jansen2019, Dymarsky2019}.  
Our results for $\Sigma^2(n,\mu)$ in Sec.\ \ref{Sec::Results} are also in 
agreement with the GOE value. 

\subsubsection{Indicators of correlations between off-diagonal matrix 
elements}\label{Sec::IndiRMT}

While the indicators of ETH given in Eqs.\ 
\eqref{Eq::VarDiag} - \eqref{Eq::SigRat} 
have 
been studied  before, this work particularly scrutinizes the presence of 
correlations between the ${\cal O}_{mn}$. To this end, we
consider the
eigenvalue distribution of the submatrices with dimension ${\cal D}'<{\cal 
D}$ around mean energy ${\bar E}$ [see Fig.\ \ref{Fig2}~(b)], and show that it 
provides a much more sensitive probe of the statistical properties of the 
${\cal O}_{mn}$. 
For a full random matrix with all matrix elements being independent, the 
eigenvalue distribution will follow the celebrated Wigner's semicircle 
\cite{dalessio2016, Mehta2004}. 
In contrast, if 
there are 
correlations 
between the ${\cal O}_{mn}$, deviations from the semicircle shape should 
emerge. Importantly, we find 
that the eigenvalue spectrum unambiguously shows that the $r_{nm}$ can not be 
represented as {\it independent} Gaussians variables, even though 
all standard indicators of the ETH are fulfilled, see Sec.\ \ref{Sec::EVs} 
below. We note that this finding is in accord with recent 
theoretical arguments from Ref.\ \cite{Dymarsky2018}. In particular,  
Ref.\ \cite{Dymarsky2018} showed that consistency with transport in a quantum 
many-body system imposes 
constraints on the matrix elements entering the ETH and requires them to be 
correlated. The strongest constraint is provided by the slowest mode 
probed by the operator ${\cal O}$. For instance, consider a system exhibiting 
diffusive transport with ${\cal O}$ being coupled to the diffusive quantity 
(such a scenario is realized in the present paper as ${\cal O} = 
S_{L/2}^z$ and spin transport is presumably diffusive in the nonintegrable 
models ${\cal H}_{1,2}$ \cite{Bertini2020}). In this case, the slowest Fourier 
mode is expected to decay as $\propto e^{-t/\tau}$ with $\tau \propto L^2/D$ 
and $D$ being the diffusion constant. While this picture would suggest that the 
${\cal O}_{mn}$ should become stuctureless and independent for frequencies 
below $\tau^{-1} \propto L^{-2}$, Ref.\ \cite{Dymarsky2018} proved that the 
scale $\Delta E_\text{RMT}$, below which genuine random-matrix may 
occur, in fact has to be parametrically smaller, $\Delta E_\text{RMT} \lesssim 
(1/\tau)/L\sim L^{-3}$. The bound $\Delta E_\text{RMT} 
\lesssim 
(1/\tau)/L$ with $\tau$ being the time scale of the 
slowest mode applies to any 
kind of transport. Therefore, in full generality, $\Delta 
E_\text{RMT} \propto 1/L^{2}$ is the loosest possible bound in a local 
system. 
Studying the signatures discussed below, we 
demonstrate in the present 
work that the scale $\Delta E_\text{RMT}$ indeed exists. Moreover, while we do 
not explicitely 
address its scaling with $L$, we specifically show that $\Delta E_\text{RMT}$ 
is drastically smaller than the scales where ``standard'' indicators of the ETH 
are already well fulfilled.

To identify the scale at which the transition to RMT behavior 
occurs, we 
analyze how the 
eigenvalue distribution depends on the width $W$ of the band, see Fig.\ 
\ref{Fig2}~(c). Specifically, let $\omega_c$ 
denote some cutoff frequency. Then we define the new operator ${\cal 
O}^{\omega_c}$ with matrix elements  
\begin{equation}\label{Eq::BandMat}
 \mathcal{O}_{mn}^{\omega_c} = \begin{cases}
                                {\cal O}_{mn}, &|E_m-E_n| < \omega_c \\
                                0, &\text{otherwise}
                               \end{cases}\ , 
\end{equation}
resulting in a band matrix with the relative bandwidth $W/{\cal D}^\prime$. 
Band random matrices have been extensively used in physics to model quantum 
systems and study their properties \cite{Casati1990, 
Fyodorov1991, Kus1991, Fyodorov1996, Borgonovi2016, Dabelow2020}. Furthermore, 
the largest eigenvalues of full (square) and band submatrices have been studied 
in 
\cite{Dymarsky2019,Dymarsky2018,Dymarsky2019_2} in connection with the 
transition from integrability to chaos as well as relaxation dynamics and 
thermalization. 
However, to the best of our knowledge, the full eigenvalue distribution of 
(band) submatrices of local operators has not been previously considered as 
a quantity to characterize the presence of 
correlations between the matrix elements ${\cal O}_{mn}$.
Provided all matrix elements are independent and 
identically distributed (except for an  overall amplitude which may depend 
on $\omega$), the eigenvalue distribution of band random matrices is 
expected to converge towards a semicircle for small $W/{\cal D}'$ 
\cite{Kus1991}, although there are
corrections at intermediate $W/{\cal 
D}'$ and the detailed shape is more complicated \cite{Molchanov}, see 
Appendix~\ref{App::RMT}.

In addition to {\it band} submatrices, we also consider the eigenvalue 
distribution of {\it full} submatrices with varying dimension ${\cal 
D}^\prime$ in Appendix \ref{App::Square}. One advantage of keeping ${\cal 
D}^\prime$ fixed and varying $W$, 
however, is that the number of eigenvalues remains unchanged and is 
comparatively large. As shown in Appendix \ref{App::Square}, the 
properties of the smaller full submatrices are in fact similar and 
consistent with our findings for the band submatrices 
\eqref{Eq::BandMat}.

Given the ordered eigenvalues $\lambda_\alpha^{\omega_c}$ obtained by 
diagonalizing  $ \mathcal{O}^{\omega_c}$ for the  cutoff frequency 
$\omega_c$, an important quantity characterizing ${\cal 
O}^{\omega_c}$ is the mean ratio $\langle r_{\omega_c} 
 \rangle$ of adjacent level spacings, 
\begin{equation}
\label{r}
 \langle r_{\omega_c}  \rangle= \frac{1}{N_r}\sum_\alpha 
\frac{\text{min}\lbrace 
\Delta_\alpha,\Delta_{\alpha+1} \rbrace}{\text{max}\lbrace 
\Delta_\alpha,\Delta_{\alpha+1} \rbrace} \ , 
\end{equation}
where $\Delta_\alpha = |\lambda_{\alpha+1}^{\omega_c} - 
\lambda_{\alpha}^{\omega_c}|$ denotes the gap between two adjacent eigenvalues 
and the averaging is performed over a number (here $N_r \approx {\cal 
D}^\prime/2$) of gaps around the center. For a 
random matrix drawn from the GOE, one expects $
r_\text{GOE}\approx 0.53$, 
while for uncorrelated Poisson distributed eigenvalues, one finds $ 
r_\text{Poisson} 
\approx 0.39$ \cite{Oganesyan2007}.  
In addition to $\langle r_{\omega_c} \rangle$, the central quantity in this 
paper is the full eigenvalue distribution 
$P_{\omega_c}(\lambda)$ of the band submatrix ${\cal O}^{\omega_c}$, 
\begin{align}\label{Eq::Distri}
 P_{\omega_c}(\lambda) &= \frac{1}{{\cal D}^\prime}\sum_{\alpha = 1}^{{\cal 
D}^\prime} 
\delta(\lambda-\lambda_{\alpha}^{\omega_c})\ ,   
\end{align}
where $\delta(\cdot)$ denotes the delta function, and it is understood that 
individual peaks are collected in small bins such that $P_{\omega_c}(\lambda)$ 
forms a ``continuous'' distribution.

Given the corrections to the semicircle 
distribution for band random matrices with intermediate $W/D'$,
a particularly simple and effective scheme to test the randomness of ${\cal 
O}^{\omega_c}$ is to compare the eigenvalue distribution $P_{\omega_c}(\lambda)$ 
with the eigenvalue distribution of the suitably randomized $\widetilde{{\cal 
O}}^{\omega_c}$. For a similar comparison of the properties of 
bare and sign-randomized matrices, see \cite{Cohen2001, Kottos2001}.
Specifically, $\widetilde{{\cal O}}^{\omega_c}$ is 
constructed by assigning random signs to the individual 
matrix elements ${\cal O}_{mn}^{\omega_c}$ (while keeping $\widetilde{{\cal 
O}}^{\omega_c}$ hermitian), 
\begin{equation}\label{Eq::TildeO}
\widetilde{{\cal O}}^{\omega_c}_{mn} = \begin{cases}
                                   {\cal O}^{\omega_c}_{mn}\ , & 50\%\ 
\text{probability} \\
(-1){\cal O}^{\omega_c}_{mn}\ , & 50\%\ 
\text{probability}
                                  \end{cases}\ . 
\end{equation}
If the matrix elements ${\cal O}^{\omega_c}_{nm}$ were random, we expect that the 
eigenvalue distribution would remain 
unchanged under this
``sign randomization''. In contrast, if the matrix elements of ${\cal 
O}^{\omega_c}$ are correlated, these correlations will be erased by 
the randomization procedure and the eigenvalue  distribution of the original and  randomized matrices 
will be different. In order to quantify the difference (and its 
dependence on $\omega_c$) between  
$P_{\omega_c}(\lambda)$ and the distribution 
$\widetilde{P}_{\omega_c}(\lambda)$ 
of the randomized operator, we introduce 
\begin{equation}
 d_2(\omega_c) = \int_{-\infty}^{\infty} [P_{\omega_c}(\lambda) - 
\widetilde{P}_{\omega_c}(\lambda)]^2\ \text{d}\lambda\ , 
\end{equation}
where $P_{\omega_c}(\lambda)$ and $\widetilde{P}_{\omega_c}(\lambda)$ should be 
understood as the continuous distributions resulting from a binning 
procedure [see below Eq.\ \eqref{Eq::Distri}].  
If $d_2(\omega_c) \to 0$, both distributions are very similar, which will
be 
interpreted as a further indication that the matrix elements ${\cal 
O}_{mn}^{\omega_c}$ are randomly distributed.

\section{Results}\label{Sec::Results}

Let us now turn to our numerical results for the matrix structure of 
$S_{L/2}^z$ in the eigenbasis of the two 
nonintegrable models ${\cal H}_1$ and ${\cal H}_2$. The properties of 
diagonal and off-diagonal matrix 
elements are discussed in Secs.\ \ref{Sec::Diag} and 
\ref{Sec::OffDiag} 
respectively, while Sec.\ \ref{Sec::EVs} presents results for the eigenvalue 
distribution of band submatrices. Additional results for the integrable 
XXZ chain can be found in Appendix~\ref{App::Int}.

\subsection{``Standard'' indicators of the ETH}\label{Sec::Standard}

\subsubsection{Diagonal matrix elements}\label{Sec::Diag}

\begin{figure}[tb]
 \centering
 \includegraphics[width=\columnwidth]{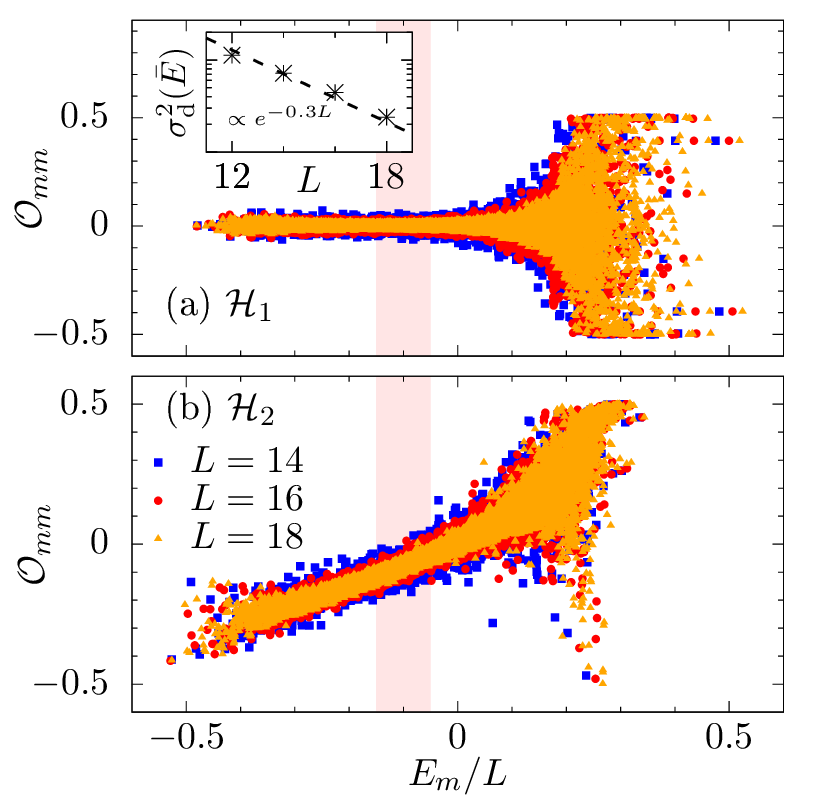}
 \caption{(Color online) Diagonal matrix elements of $S_{L/2}^z$ in the 
eigenbasis of (a) ${\cal H}_1$ and (b) ${\cal H}_2$, for system sizes $L = 
14,16,18$. The shaded area indicates the energy window $E_m/L \in 
[-0.15,-0.05]$ which is used in the 
following to further study the properties of off-diagonal matrix elements. For 
the example of ${\cal H}_1$, the 
inset in (a) shows that the variance $\sigma_\text{d}^2(\bar{E})$ of the 
${\cal O}_{mm}$ in this window decreases exponentially with increasing 
$L$.}
 \label{Fig3}
\end{figure}

As a first step, we study the diagonal part of the ETH. 
To this end, Figs.\ \ref{Fig3}~(a) and (b) show the matrix 
elements ${\cal O}_{mm} = \bra{m} S_{L/2}^z\ket{m}$
as a function of the corresponding energy 
density $E_m/L$ for ${\cal H}_{1,2}$ and different system 
sizes $L = 14,16,18$. 
For energy densities in the center of the 
spectrum, we find that the ``cloud'' of matrix elements becomes narrower with 
increasing $L$ \cite{steinigeweg2013, beugeling2014, Brenes2020_1}. This 
finding is in good accord with the ETH prediction that the 
${\cal O}_{mm}$ should form a ``smooth'' function of energy in the 
thermodynamic 
limit $L \to \infty$. 
At the edges of the spectrum, the scaling with $L$ is 
significantly slower. Especially for ${\cal 
H}_1$ [Fig.\ \ref{Fig3}~(a)] and $E_m/L \gtrsim 0.2$, the ${\cal 
O}_{mm}$ are found to fluctuate very strongly. This can be 
understood as follows. The 
eigenstates of ${\cal H}_1$ with the highest energies are 
weakly dressed domain-wall states. Consider, for instance, the states 
$\ket{n_1} 
= 
\ket{\uparrow \cdots \uparrow \downarrow \cdots \downarrow}$ and $\ket{n_2} = 
\ket{\downarrow \cdots \downarrow \uparrow \cdots \uparrow}$ (note that   
$\ket{n_1}$ and $\ket{n_2}$ are not exact eigenstates). While $\ket{n_1}$ and 
$\ket{n_2}$ have almost the same energy, one finds that 
$\bra{n_1}S_{L/2}^z\ket{n_1} \approx 1/2$ whereas $\bra{n_2}S_{L/2}^z\ket{n_2} \approx 
-1/2$ such that ETH is not satisfied. We expect the range of energy densities  
where the ETH applies to increase with $L$.  

Given the distribution of the ${\cal O}_{mm}$, we 
restrict ourselves in the following to eigenstates in an energy 
window $E_m/L \in [-0.15,-0.05]$ which is close to the center of the 
spectrum (shaded area in Fig.\ \ref{Fig3}). As 
shown in the inset of Fig.\ \ref{Fig3}~(a), 
the variance $\sigma_\text{d}^2(\bar{E})$ of the ${\cal O}_{mm}$ in this 
window decays approximately 
exponentially with $L$ (at least for the system sizes numerically available), 
indicating that the diagonal part of the ETH is fulfilled.

\subsubsection{Off-diagonal matrix elements}\label{Sec::OffDiag}

\begin{figure}[tb]
 \centering
 \includegraphics[width=\columnwidth]{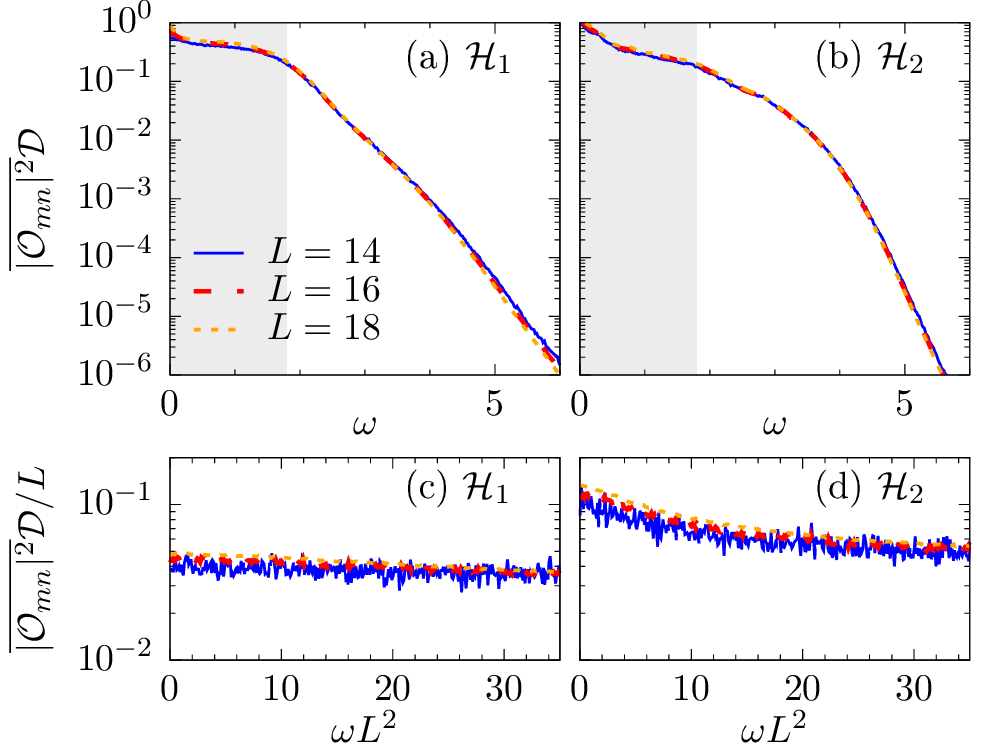}
 \caption{(Color online) [(a),(b)] Running averages 
$\overline{|{\cal O}_{mn}|^2}$ of matrix elements in the energy window 
$\bar{E}/L 
\in [-0.15,-0.05$], calculated for system sizes $L = 14,16,18$ and 
frequency bins of width $\Delta \omega = 0.01$. [(c),(d)] Close-up of the 
low-frequency regime using a bin width of $\Delta \omega = 5\times 
10^{-4}$. Note that both the 
horizonal and the vertical axis have been rescaled.
Panels (a) and (c) show results for ${\cal H}_1$, while (b) and 
(d) show data for ${\cal H}_2$. The shaded area in panels (a) and (b) indicates 
the $\omega$ range which is probed when considering a square-shaped 
submatrix with eigenstates in an energy interval of width $\Delta E/L = 0.1$.}
 \label{Fig4}
\end{figure}

Let us now analyze the properties of the off-diagonal matrix elements  
${\cal O}_{mn} = \bra{m} S_{L/2}^z \ket{n}$. 
In view of the previous results for the diagonal matrix elements in Fig.\ 
\ref{Fig1}, we focus on eigenstates with mean energy density 
$\bar{E}/L \in [-0.15,-0.05]$ as said above. 
For matrix elements in this window, running averages $\overline{|{\cal 
O}_{mn}|^2}$ of their absolute squares  
are shown in Figs.\ \ref{Fig4}~(a) and (b) 
as a function of $\omega$ both for ${\cal H}_1$ and 
${\cal H}_2$. The data are obtained for frequency intervals of width $\Delta 
\omega = 10^{-2}$ and system sizes $L = 14,16,18$. Overall, the 
situation is qualitatively similar for the two models ${\cal H}_{1,2}$. Namely, 
$\overline{|{\cal 
O}_{mn}|^2}$ decays comparatively slowly at low 
frequencies, while a substantially quicker (presumably superexponential 
\cite{Abanin2015, Avdoshkin2019}) decay 
can be 
found at higher $\omega$.
In Figs.\ 
\ref{Fig4}~(a) and (b) the values of $\overline{|{\cal 
O}_{mn}|^2}$ for different $L$ form smooth curve which collapse on each 
other when rescaled by the 
respective Hilbert-space dimension ${\cal D}$. [The rescaling 
by ${\cal D}$ accounts for the factor $\Omega^{-\tfrac{1}{2}}(\bar{E})$ 
in Eq.\ \eqref{Eq::ETH}.] Except for a 
prefactor, these smooth curves correspond to the function $f^2(\bar{E},\omega)$ 
from the ETH ansatz~\eqref{Eq::ETH}. 
\begin{figure}[tb]
 \centering
 \includegraphics[width=\columnwidth]{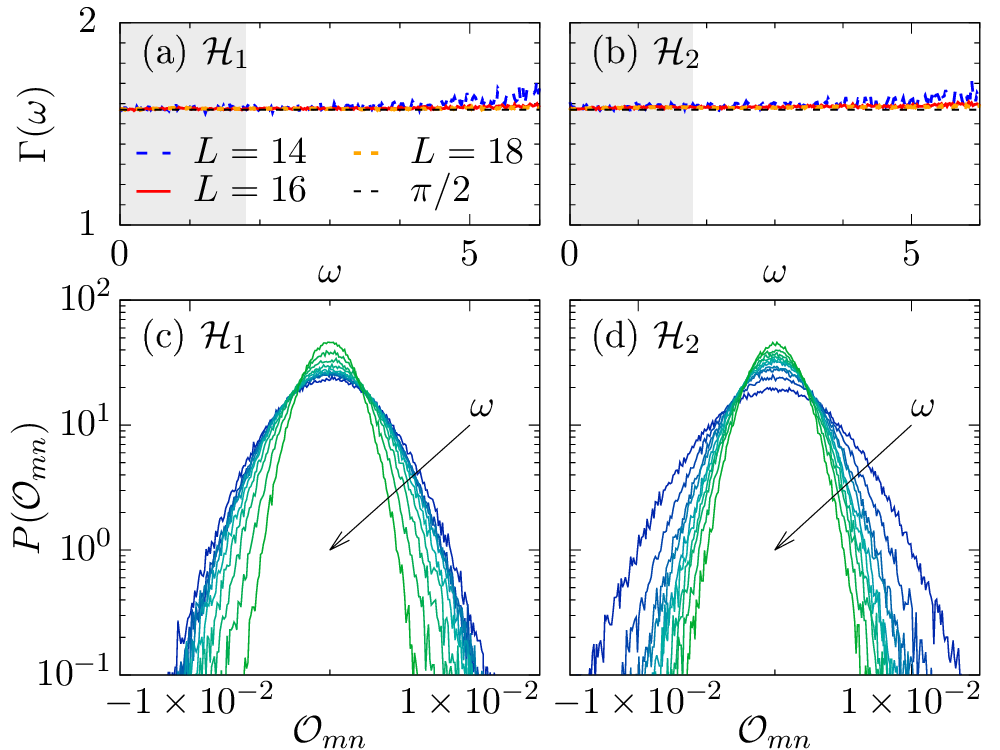}
 \caption{(Color online)  [(a),(b)] $\Gamma(\omega)$ for matrix elements in 
the energy window 
$\bar{E}/L \in [-0.15,-0.05$] and 
system sizes $L = 14,16,18$. The dashed line indicates the value $\pi/2$ for a 
Gaussian distribution. The shaded area indicates 
the $\omega$ range which is probed when considering a square-shaped 
submatrix with eigenstates in an energy interval of width $\Delta E/L = 0.1$.
[(c),(d)] Distribution $P({\cal O}_{mn})$ of off-diagonal 
matrix elements for $L = 
18$ and $\omega = 0.2,0.4,\dots,2$ (arrow). The data are 
collected in 
frequency bins $[\omega-\Delta \omega/2,\omega+\Delta \omega/2]$ with $\Delta 
\omega = 0.002$. (a) and (c) show results for 
${\cal H}_1$, while (b) and 
(d) show data for ${\cal H}_2$.}
 \label{Fig5}
\end{figure}

For a more detailed analysis of $\overline{|{\cal 
O}_{mn}|^2}$, Figs.\ \ref{Fig4}~(c) and (d) show a close-up 
of the low-frequency regime (note that the horizontal and the vertical 
axis have been rescaled to account for possible finite-size 
effects at such small $\omega$). For both ${\cal H}_1$ and ${\cal H}_2$, we 
observe
that $\overline{|{\cal 
O}_{mn}|^2}$ clearly approaches a nonzero value as $\omega \to 0$ with an 
approximately constant plateau for small $\omega L^2$, where 
the data collapse achieved by the $L^2$-rescaling indicates diffusive spin 
dynamics
(see also the 
discussion in Ref.\ \cite{Brenes2020_2}). 

Next, in order to study the distribution of the ${\cal O}_{mn}$, Figs.\ 
\ref{Fig5}~(a) and (b) show the 
frequency-dependent ratio 
$\Gamma(\omega)$, defined in \ Eq.\ \eqref{Eq::Gamma}. For small 
$\omega$, we find that $\Gamma(\omega)$ is close to the 
Gaussian value $\pi/2$, while visible deviations appear at higher 
frequencies. However, these 
deviations  decrease with the increasing  system size $L$, 
indicating that the ${\cal O}_{mn}$ follow a Gaussian distribution over a 
wide range of 
frequencies if $L$ is sufficiently large.
In addition, the full distribution $P({\cal O}_{mn})$ of the off-diagonal 
matrix elements is shown in 
Figs.\ 
\ref{Fig5}~(c) and (d) for system size $L = 18$ and 
frequencies $\omega = 0.2,0.4,\dots,2$. For all curves shown, we find that 
$P({\cal O}_{mn})$ is 
indeed well described by the Gaussians with zero mean (see also Refs.\ 
\cite{beugeling2015, LeBlond2019, Luitz2016}). The 
width of the Gaussians is found to decrease with increasing $\omega$, which is 
consistent with the data for $\overline{|{\cal 
O}_{mn}|^2}$ shown in Figs.\ \ref{Fig4}~(a) and (b).

\begin{figure}[tb]
 \centering
 \includegraphics[width=\columnwidth]{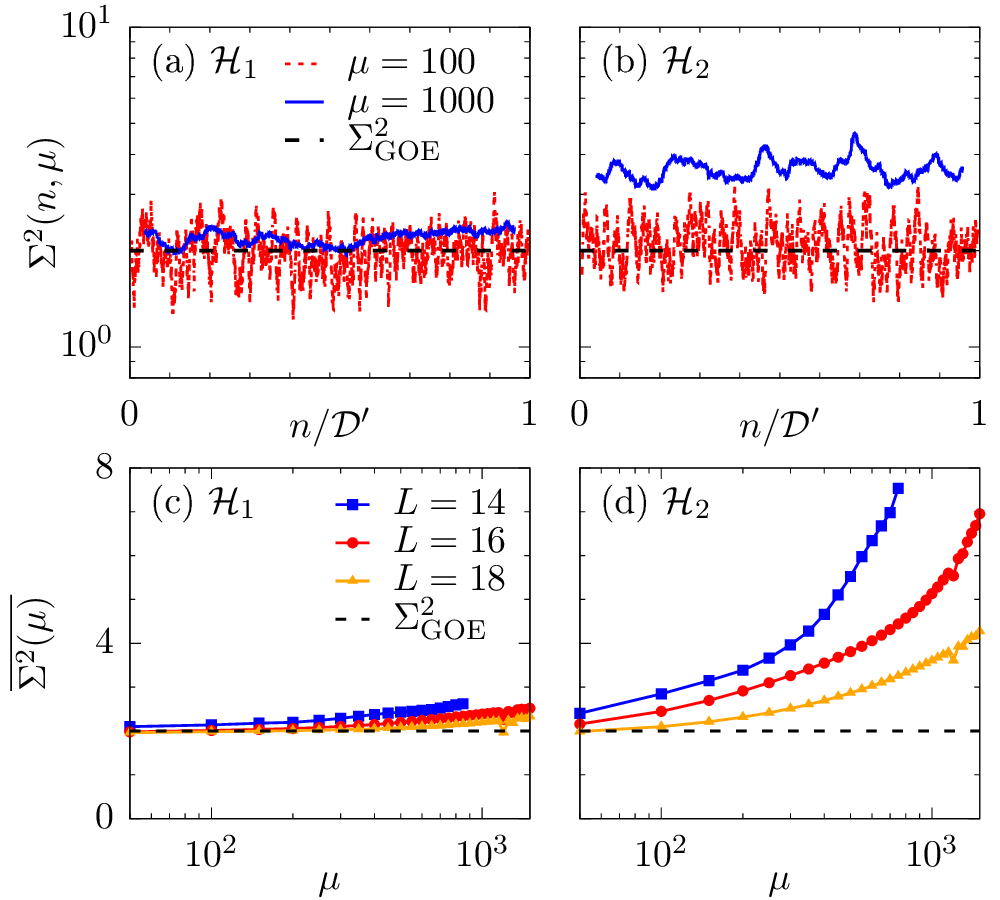}
 \caption{(Color online) [(a),(b)] Ratio $\Sigma^2(n,\mu)$ between the 
variances of diagonal and off-diagonal matrix elements for two different 
square sizes $\mu = 100,1000$ and all embeddings $n \in [1+\mu/2,{\cal 
D}'-\mu/2]$. The data are obtained for system size $L = 18$ and the dashed line 
indicated the value $\Sigma_\text{GOE}^2= 2$ predicted from RMT. [(c),(d)] 
Average value 
$\overline{\Sigma^2(\mu)}$ versus $\mu$ for system sizes $L = 14,16,18$. Note 
that for the largest system size $L = 18$, the maximum $\mu = 2\times 10^3$ 
shown here is still considerably smaller than the full submatrix dimension 
${\cal D}^\prime \approx 1.3\times 10^4$. 
Panels (a) and (c) show data for ${\cal H}_1$ while (b) and (d) show data for 
${\cal H}_2$.} 
 \label{Fig6}
\end{figure}

Let us finally comment on the shaded gray area at low frequencies in Figs.\ 
\ref{Fig4}~(a), (b) and Figs.\ \ref{Fig5}~(a), (b). This area indicates the 
frequency range 
which is covered when considering a square-shaped submatrix in the interval 
$\bar{E}/L \in [-0.15,-0.05]$, cf.\ Fig.\ \ref{Fig2}. Specifically, 
since this interval has a width $\Delta E/L = 0.1$, the largest energy 
difference for $L = 18$ is $\omega_\text{max} = 0.1 L = 1.8$. 
Therefore, when studying the eigenvalue distribution of such a submatrix further 
below, we are probing the region where $\overline{|{\cal 
O}_{mn}|^2}$ varies comparatively slowly and $\Gamma(\omega) \approx \pi/2$.

To conclude the analysis of the off-diagonal matrix elements, Figs.\ 
\ref{Fig6}~(a) and (b) show the ratio $\Sigma^2(n,\mu)$ between the variances 
of 
diagonal and off-diagonal matrix elements. Specifically, the data are obtained 
for $L = 18$ with two 
different square sizes $\mu = 100,1000$ and all possible embeddings along the 
diagonal of the submatrix with dimension ${\cal D}^\prime$. (Note that for the 
chosen energy window, we have ${\cal D}^\prime \approx {\cal D}/4$.)
For small $\mu = 
100$, we find that $\Sigma^2(n,\mu)$ fluctuates around the 
GOE value $\Sigma_\text{GOE}^2 = 2$, both for ${\cal H}_1$ and ${\cal 
H}_2$. Increasing the square size to $\mu = 
1000$, we observe that the fluctuations of $\Sigma^2(n,\mu)$ are visibly 
reduced. For the larger value of $\mu$, $\Sigma^2(n,\mu)$ is still rather 
close to 
$\Sigma_\text{GOE}^2$ for ${\cal 
H}_1$, while clear deviations between 
$\Sigma^2(n,\mu)$ and $\Sigma_\text{GOE}^2$ can be seen in the case of 
${\cal H}_2$.

Figures \ref{Fig6}~(c) and (d) show 
the averaged 
value, 
\begin{equation}
 \overline{\Sigma^2(\mu)} = \frac{1}{{\cal D}'-\mu}\sum_{n=1+\mu/2}^{{\cal 
D}'-\mu/2} \Sigma^2(n,\mu)\ , 
\end{equation}
calculated from all embeddings $n \in [1+\mu/2,{\cal 
D}'-\mu/2]$.
We find that $\overline{\Sigma^2(\mu)} \approx \Sigma_\text{GOE}^2$ for small 
$\mu$, while $\overline{\Sigma^2(\mu)}$ monotonously grows with increasing 
$\mu$ (this growth is particularly pronounced in the case of ${\cal 
H}_2$). This behavior of $\overline{\Sigma^2(\mu)}$ follows from
 the $\omega$ dependence of 
$\overline{|{\cal 
O}_{mn}|^2}$ discussed in Fig.\ \ref{Fig4}. Since 
$\overline{|{\cal 
O}_{mn}|^2}$ decreases with increasing $\omega$, the variance 
$\sigma_\text{od}^2(n,\mu)$ likewise decreases with increasing $\mu$, simply because matrix 
elements at higher frequencies are included.
Comparing $\overline{\Sigma^2(\mu)}$ for different system sizes, 
we find that $\overline{\Sigma^2(\mu)}$ remains closer to $\Sigma_\text{GOE}^2$ 
for a wider range of $\mu$ as $L$  increases (see also 
\cite{jansen2019}). 
Therefore in the thermodynamic limit $L \to \infty$ one can expect  ${\cal 
O}_{nm}$ to approach an actual random matrix drawn from the GOE, at least 
for a finite region around the diagonal. 

To summarize, in this subsection we considered different quantities 
conventionally 
considered as standard indicators of ETH. 
The results presented in Figs.\ \ref{Fig3} - \ref{Fig6} confirm 
that the matrix structure of the local spin-$1/2$ operator ${\cal O} = 
S_{L/2}^z$ 
in the eigenbasis of the nonintegrable models ${\cal H}_{1,2}$ is 
in good agreement with the ETH ansatz \eqref{Eq::ETH}, at least for the chosen energy 
window close to the center of the spectrum. 
Nevertheless, in the next subsection, we will show that the matrix elements 
${\cal 
O}_{mn}$ within this energy window can not be considered as fully 
uncorrelated, i.e., the standard indicators in Figs.\ \ref{Fig3} - 
\ref{Fig6} are not sufficient when it comes to the statistical properties of  
the ${\cal 
O}_{mn}$. 

\subsection{Beyond ``standard'' indicators: Eigenvalue 
distribution of band submatrices}\label{Sec::EVs}

\begin{figure}[tb]
 \centering
 \includegraphics[width=\columnwidth]{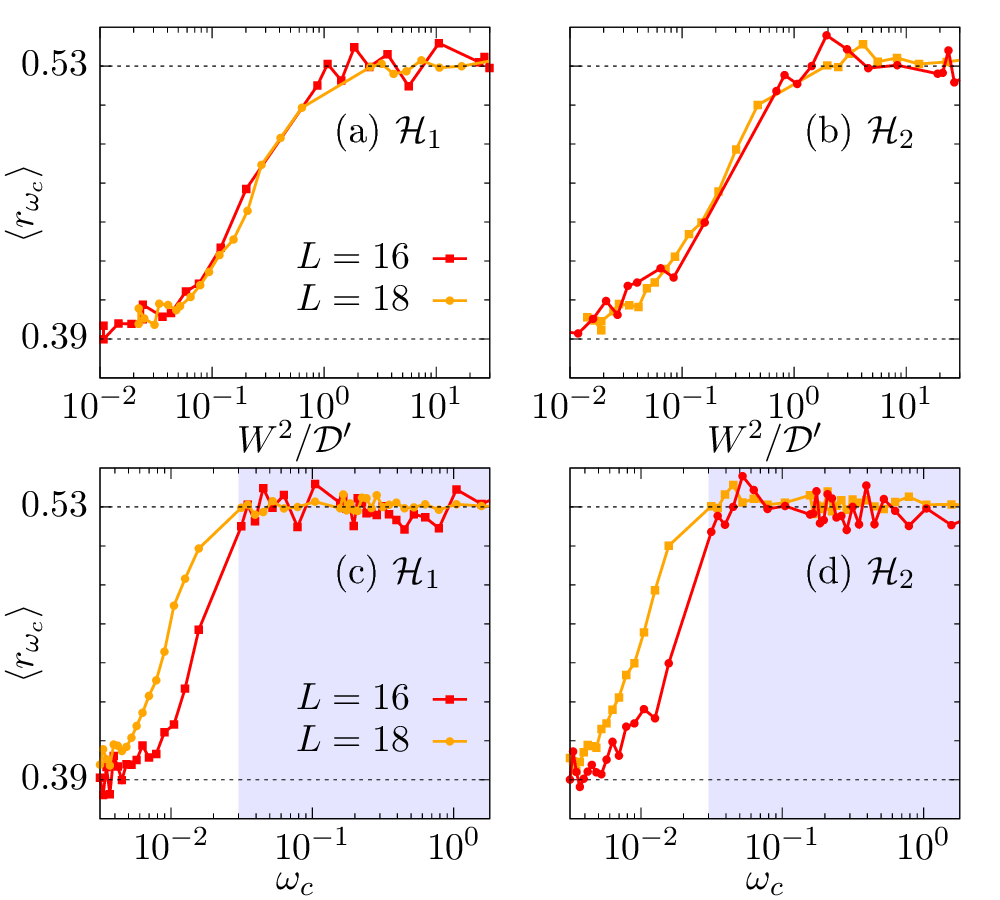}
 \caption{(Color online) Mean ratio $\langle r_{\omega_c} \rangle$ of adjacent 
level spacing of the operator ${\cal O}^{\omega_c}$ versus [(a),(b)] the 
scaling parameter $W^2/{\cal D}'$, and [(c),(d)] the cutoff frequency 
$\omega_c$. Panels (a) and (c) show data for ${\cal H}_1$ while (b) and (d) 
show data for ${\cal H}_2$. The data is obtained for system sizes $L = 16, 
18$ and the dashed horizontal lines indicate the GOE value 
$r_\text{GOE} \approx 0.53$ and the Poisson value $
r_\text{Poisson} \approx 0.39$. For our analysis of 
$P_{\omega_c}(\lambda)$, we restrict ourselves to $\omega_c 
\gtrsim 0.03$ [shaded area in (c) and (d)], such that $\langle 
r_{\omega_c} \rangle \approx  r_\text{GOE} $.}
 \label{Fig7}
\end{figure}

We now turn to the eigenvalue distribution for the band 
submatrices centered around $\bar{E}/L = -0.1$ with $\Delta E/L = 0.1$.  
First we discuss the ratio of the adjacent level spacings
$\langle r_{\omega_c} \rangle$ defined in \eqref{r}, which is shown in Fig.\ 
\ref{Fig7} versus $W^2/{\cal D}^\prime$ [panels (a) and (b)] as well as 
versus $\omega_c$ [panels (c) and (d)].
We find that the 
behavior of $\langle 
r_{\omega_c} \rangle$ is very similar for ${\cal H}_1$ and ${\cal H}_2$. 
Specifically, over a wide range of $\omega_c$, $\langle r_{\omega_c} \rangle 
\approx 0.53$ approximately matches the 
GOE value, while the transition towards 
the Poissonian value $\langle r_{\omega_c} \rangle \approx 0.39$ occurs when the 
bandwidth becomes too narrow. 
The crossover from $r_\text{GOE}$ to $r_\text{Poisson}$ can be understood from 
the well-known fact that the eigenstates of a band random matrix with a 
sufficiently small value of $W^2/D'$ are localized and the eigenvalues become 
uncorrelated \cite{Fyodorov1991}. 
In order to avoid localization effects while studying the eigenvalue 
distribution $P_{\omega_c}(\lambda)$, we restrict our analysis to $\omega_c 
\gtrsim 0.03$ (shaded area in Fig.\ \ref{Fig7}), such that $\langle 
r_{\omega_c} \rangle \approx r_\text{GOE}$. 

\begin{figure}[tb]
 \centering
 \includegraphics[width=\columnwidth]{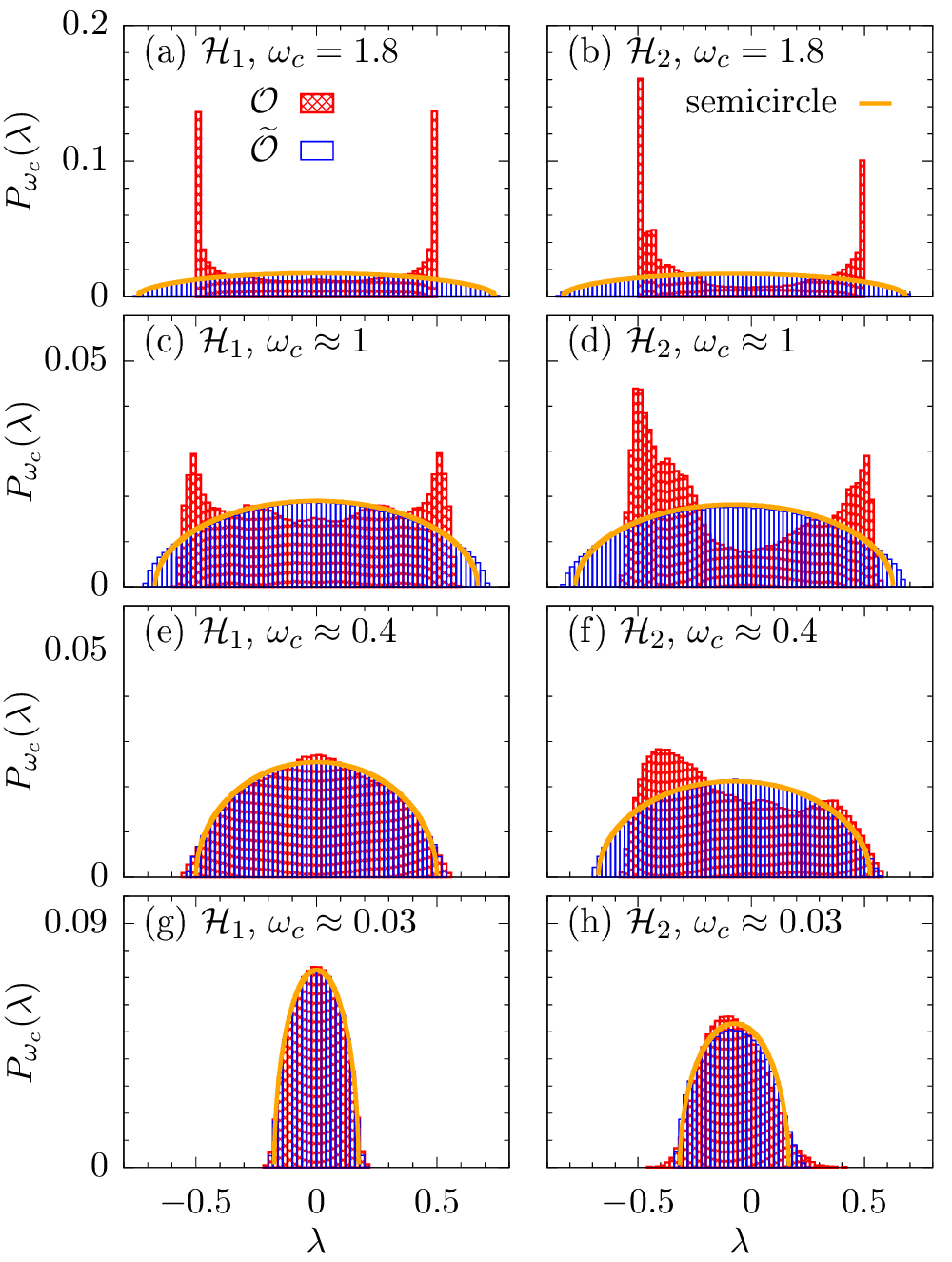}
 \caption{(Color online) Eigenvalue distributions $P_{\omega_c}(\lambda)$ and 
$\widetilde{P}_{\omega_c}(\lambda)$ of bare and sign-randomized 
submatrices in the energy window $\bar{E}/L \in [-0.15,-0.05]$ for system size 
$L = 18$. The cutoff frequencies are chosen as [(a),(b)] $\omega_c = 1.8$  
(i.e.\ the full nonband
submatrix of dimension ${\cal D}' < {\cal D}$), 
[(c),(d)] $\omega_c \approx 1$, [(e),(f)] $\omega_c  \approx 0.4$, and 
[(g),(h)] $\omega_c  \approx 0.03$. For 
comparison, the solid curves indicate a semicircle distribution. Left 
column shows data for ${\cal H}_1$, while right column shows data for 
${\cal H}_2$. The skewed distribution in the case of ${\cal H}_2$ 
can be explained by the diagonal matrix elements 
${\cal O}_{mm}$ [see Fig.\ \ref{Fig3}~(b)] which have a mean that (i) is 
nonzero within the energy window and (ii) grows with $E$, in contrast to the 
case of ${\cal H}_1$, cf. Fig.\ \ref{Fig3}~(a).}
 \label{Fig8}
\end{figure}

In Fig.\ \ref{Fig8}, we show the full spectrum 
$P_{\omega_c}(\lambda)$ for ${\cal H}_{1,2}$ with $L = 18$ and four 
exemplary choices of the cutoff frequency $\omega_c$. Specifically, we have 
chosen $\omega_c = 1.8$ (i.e.\ the full nonband submatrix), as well as $\omega_c 
\approx 1$, $\omega_c \approx 0.4$, 
and $\omega_c \approx 0.03$, which are all above the transition point of 
$\langle r_{\omega_c} \rangle$. 
In all cases, we compare the spectrum of the bare operator 
${\cal O}^{\omega_c}$ to the distribution 
$\widetilde{P}_{\omega_c}(\lambda)$ of the sign-randomized version 
$\widetilde{{\cal O}}^{\omega_c}$, see Eq.\ \eqref{Eq::TildeO}.
As can be clearly seen in Figs.\ \ref{Fig8}~(a) and (b), 
$P_{\omega_c}(\lambda)$ and $\widetilde{P}_{\omega_c}(\lambda)$ 
differ strongly for the largest $\omega_c$ considered. Specifically, 
while $\widetilde{P}_{\omega_c}(\lambda)$ closely follows the semicircle 
law appropriate for random matrices, $P_{\omega_c}(\lambda)$ 
still exhibits pronounced peaks at $\pm 1/2$, which is reminiscent to the 
original spectrum of the  
spin-$1/2$ operator.
This deviation between 
$P_{\omega_c}(\lambda)$ and $\widetilde{P}_{\omega_c}(\lambda)$ illustrates 
that the matrix elements ${\cal O}_{mn}$ of a small submatrix cannot 
automatically be considered as independent 
random 
variables, notwithstanding  all standard indicators of ETH being in agreement 
with the Gaussian distribution. This is a main result of the present 
paper.  

Lowering the cutoff frequency to $\omega_c \approx 1, 0.4$ and $\omega_c 
\approx 0.03$ 
[see Figs.\ \ref{Fig8}~(c)-(h)], we find that the spectra of the 
bare and the randomized submatrices become more and more similar. Especially  
for 
${\cal H}_1$, $P_{\omega_c}(\lambda)$ and 
$\widetilde{P}_{\omega_c}(\lambda)$ are very similar for $\omega_c \approx 0.4$ 
and virtually indistinguishable from each 
other for $\omega_c \approx 0.03$. Moreover, the bulk of the spectrum is 
convincingly described by a 
semicircular distribution (the width of the semicircle shrinks with $\omega_c$), 
while small deviations from a perfect semicircle 
can be 
observed at the spectral edges. This similarity of  
$\widetilde{P}_{\omega_c}(\lambda)$ and $P_{\omega_c}(\lambda)$ as well as 
their 
semicircular shape can be interpreted as an indication that the correlations 
between the matrix 
elements are significantly reduced for frequencies 
around and below  $\omega \lesssim \Delta E_{\rm RMT}\approx 0.4$, i.e., on 
these 
smaller scales the ${\cal 
O}_{mn}$ can be represented as  independent random variables. 
This is another central result of the present work. 
Let us emphasize that the full distribution 
$P_{\omega_c}(\lambda)$ is sensitive to the RMT scale $\Delta E_{\rm RMT}$ 
while the mean gap ratio $\langle r_{\omega_c} \rangle$ takes on the GOE value 
for 
all $\omega_c$ considered in Fig.~\ref{Fig8}.

\begin{figure}[tb]
 \centering
 \includegraphics[width=\columnwidth]{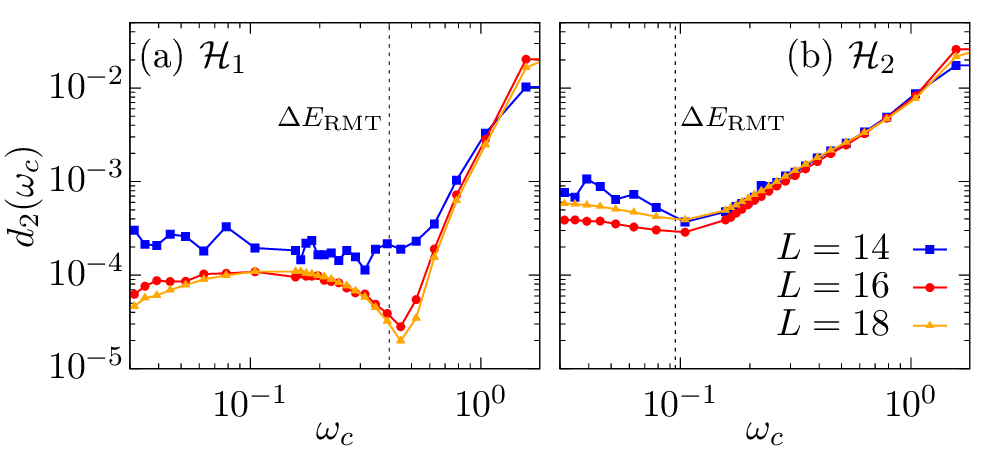}
 \caption{(Color online) $d_2(\omega_c)$ for (a) ${\cal H}_1$, and (b) ${\cal 
H}_2$. System sizes are chosen as $L = 14,16,18$. The dashed vertical lines 
indicate the approximate 
location of $\Delta E_\text{RMT}$ below which random-matrix behavior occurs.}
 \label{Fig9}
\end{figure}

Comparing properties of ${\cal H}_1$ and ${\cal H}_2$, we find that 
$P_{\omega_c}(\lambda)$ and $\widetilde{P}_{\omega_c}(\lambda)$   still 
differ visibly for $\omega_c \approx 0.4$ in the case of ${\cal H}_2$, see
Fig.\ \ref{Fig8}~(f), but become very similar for the smaller 
$\omega_c \approx 0.03$, see Fig.\ \ref{Fig8}~(h). This is in accord with 
Fig.\ \ref{Fig9} below, which suggests that 
$\Delta 
E_{\rm RMT} \approx 0.1$ is smaller in the case of ${\cal H}_2$.

While it certainly would be desirable to study the eigenvalue distribution 
$P_{\omega_c}(\lambda)$ for even smaller values of ${\omega_c}$ in a controlled 
manner, this is difficult to do numerically as the value of $W$ and the 
relative 
bandwidth size $W/{\cal D}'$ become too small. Likewise, if one instead 
diagonalizes full nonband submatrices with smaller dimension ${\cal 
D}^\prime$, see Appendix \ref{App::Square}, the number of eigenvalues becomes 
significantly reduced, which complicates the analysis.

Finally, Figs.\ \ref{Fig9}~(c) and (d) show the difference $d_2(\omega_c)$ 
between the two distributions $P_{\omega_c}(\lambda)$ and
$\widetilde{P}_{\omega_c}(\lambda)$. Consistent with our previous observation 
in Fig.\ \ref{Fig8}, we find that $d_2(\omega_c)$ decreases upon reducing 
$\omega_c$. Moreover, the decrease is slower in the case of ${\cal H}_2$. The 
minimum of $d_2(\omega_c)$ is reached around the values of $\omega_c$ which we 
associate with $\Delta E_{\rm RMT}$ ($\Delta E_{\rm RMT} \approx 0.4$ in 
case of ${\cal H}_1$ and $\Delta E_{\rm RMT} \approx 0.1$ in case of ${\cal 
H}_2$), and $d_2(\omega_c)$ remains low for 
smaller~$\omega_c$.

\section{Discussion}\label{Sec::Summary}

In this paper we have studied matrix elements of the spin-$1/2$ operator 
${\cal O} = S_{L/2}^z$ in the eigenbasis of two nonintegrable 
quantum spin chains. Specifically, we have considered the one-dimensional XXZ 
model in the 
presence of two different integrability-breaking perturbations: an 
additional next-nearest neighbor interaction and a single-site magnetic field 
in the center.  
For these models and an energy window close to the center of the spectrum, we 
have shown that the matrix elements of ${\cal O}$ are in good agreement with 
the eigenstate thermalization hypothesis ansatz in the sense that (i) variance of 
the diagonal matrix elements decreases exponentially with the increasing system 
size, (ii) the off-diagonal matrix elements follow a Gaussian distribution with 
a variance that depends smoothly on the energy difference $\omega$, and (iii) 
the ratio between the variances of diagonal and off-diagonal matrix elements 
approximately takes on the value predicted by random matrix theory.
Overall, our results are in full agreement with previous works 
\cite{Brenes2020_1, Brenes2020_2, Pandey2020, Santos2020} and the conventional 
expectation that for local operators and nonintegrable Hamiltonians
the ETH is satisfied.

The central question of this paper was to study to what extent 
off-diagonal matrix elements ${\cal O}_{mn}$ can be treated as 
independently drawn random variables. To this end, we have considered 
submatrices around a fixed mean energy $\bar{E}$ and restricted the 
${\cal O}_{mn}$ to lie inside a sufficiently narrow band $|E_n-E_m|\leq 
\omega_c$. We have established the form of the full eigenvalue distribution to 
be a sensitive 
probe to correlations between matrix elements. By comparing 
the eigenvalue distribution of the band submatrix with its 
sign-randomized counterpart \eqref{Eq::TildeO}, 
we have shown that the ${\cal O}_{mn}$ cannot be considered as independently 
distributed, even 
on scales where ${\cal O}_{mn}$ follow a Gaussian distribution with a 
variance that varies comparatively slowly with $\omega$, i.e., on the scales 
where the ETH 
function $f(\bar{E},\omega)$ is approximately constant.  
Specifically, while the spectrum of the 
sign-randomized matrix closely 
followed the semicircle law, matching the theoretical expectation for a random 
matrix, the 
eigenvalue distribution of the original submatrix was found to exhibit 
signatures 
of the spin operator, implying
correlations between the matrix elements. When the cutoff frequency $\omega_c$ 
is sufficiently reduced, we have found that the
eigenvalue distribution of the original and the sign-randomized operator 
become similar and well described by a semicircle. 
The energy scale $\Delta E_{\rm RMT}$ when this transition occurs marks the 
onset of validity of the random-matrix behavior. 

It should be noted that many important results rooted in the ETH are not 
sensitive to the statistics of the off-diagonal matrix elements ${\cal 
O}_{mn}$. This includes the central argument that the ETH ensures 
thermalization \cite{rigol2005, Rigol2012, dalessio2016}, which essentially 
relies on the exponential smallness of the ${\cal O}_{mn}$.
At the same time, within the contemporary understanding of 
ETH, it is often assumed that matrix elements posses additional statistical 
properties 
matching the GOE (or some other appropriate Gaussian ensemble), see 
e.g.~\cite{beugeling2015,dalessio2016,Luitz2016,mondaini2017}.
In this work, we have 
advocated that below a certain energy scale all statistical 
properties of the 
off-diagonal matrix elements would match 
those of a Gaussian random matrix. We have provided numerical evidence that 
this onset of random-matrix behavior takes place below a certain energy 
scale $\Delta E_\text{RMT}$ for specific models and 
observables. Our results suggest that for frequencies $\omega < \Delta 
E_\text{RMT}$, the notion of \mbox{(pseudo-)}randomness of the $r_{mn}$ 
entering the 
ETH can be interpreted in an even stricter sense.   
At the same time, we have clearly seen that  
the scale $\Delta E_\text{RMT}$, where this transition to genuine random-matrix 
behavior occurs, is distinctly 
smaller than the scales on which ``standard'' indicators of the ETH are 
fulfilled. 

Our work raises a number of straightforward questions. First, we note that our 
numerical observation $\Delta E_{\rm 
RMT} \ll E_\tau$ mirrors the analytical bound $\Delta E_{\rm RMT}  
\lesssim E_\tau/L$ established in \cite{Dymarsky2018}, where 
$E_\tau$ 
is defined as the width of the plateau of 
$f(\bar{E},\omega)$ (note that $E_\tau$ is sometimes referred to as Thouless 
energy 
\cite{Serbyn2017}). A natural question would be to establish 
the scaling of $\Delta E_{\rm RMT}$ with the system size $L$ and, 
in particular, to investigate if $(\Delta E_{\rm RMT})^{-1}$ can be associated 
with the time scale of late time chaos at which the dynamics of 
various observables is captured by RMT \cite{Cotler2017, Cotler2019, 
Moudgalya2019, Schiulaz2019}. 

Another direction is to contrast the behavior in   
chaotic systems with the integrable counterparts. We repeat the analysis of 
section \ref{Sec::Results} for the integrable XXZ model in the Appendix 
\ref{App::Int}. One particular observation to point out is that 
off-diagonal matrix elements in the integrable case also can be regarded as 
random and independent, although not Gaussian, below a certain 
energy scale. 
We leave for the future the question of better understanding this transition, 
and the universal properties of ${\cal O}_{mn}$ in the integrable case.

Eventually, one avenue of research is to characterize the nature of the 
correlations between off-diagonal matrix elements for $\omega > \Delta 
E_\text{RMT}$, and to understand their potential impact on self-averaging 
properties of the ${\cal O}_{mn}$ exploited in various works \cite{Richter2019, 
Richter2020, Dabelow2020, Nation2019}.   
At the same time, it would be interesting to study the connection between 
universal properties of ${\cal O}_{mn}$ at small frequencies and transport, 
which could be diffusive or ballistic \cite{Bertini2020}.

\subsection*{Acknowledgements}

We thank M. Bergfeld and M. Lamann for discussions and helpful comments on the 
manuscript. This work has been funded by the Deutsche Forschungsgemeinschaft 
(DFG) - 
Grants No.\ 397107022 (GE 1657/3-1), No.\ 397067869 (STE 
2243/3-1), No.\ 355031190 - within the DFG Research Unit FOR 2692. 
A.\ D.\ acknowledges support of the Russian Science Foundation (Project No. 17-12-01587).

\appendix

\section{Eigenvalue distribution of full nonband
submatrices}\label{App::Square}

\begin{figure}[tb]
 \centering
 \includegraphics[width=\columnwidth]{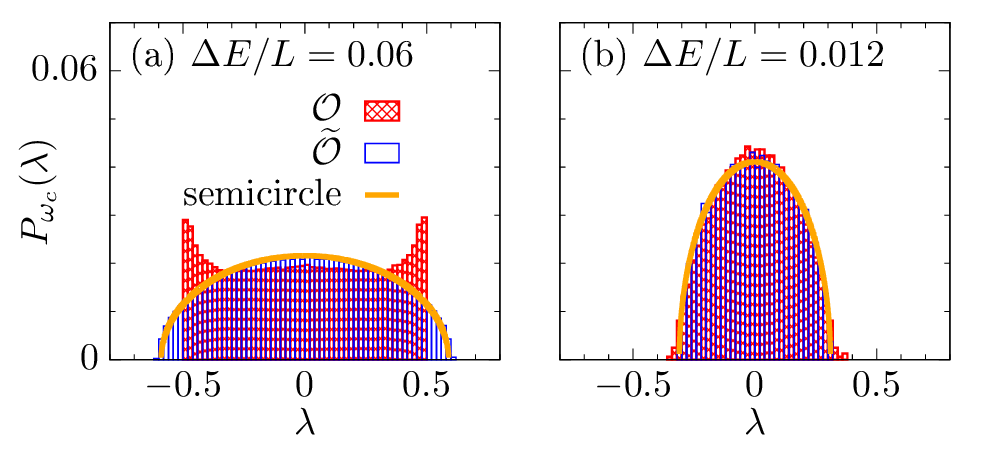}
 \caption{(Color online) Eigenvalue 
distributions $P_{\omega_c}(\lambda)$ and 
$\widetilde{P}_{\omega_c}(\lambda)$ for full (nonband) submatrices in the case 
of 
the model ${\cal H}_1$ with $L = 18$. The matrices are centered around the mean 
energy $\bar{E}/L = 
-0.1$, but the width of the energy window is now smaller compared to the main 
text, i.e., we choose 
(a) $\Delta E/L = 0.06$ and (b) $\Delta E/L = 0.012$, which would correspond 
cutoff frequencies $\omega_c \approx 1$ and $\omega_c \approx 0.2$.
}
 \label{Fig10}
\end{figure}
\begin{figure}[tb]
 \centering
 \includegraphics[width=\columnwidth]{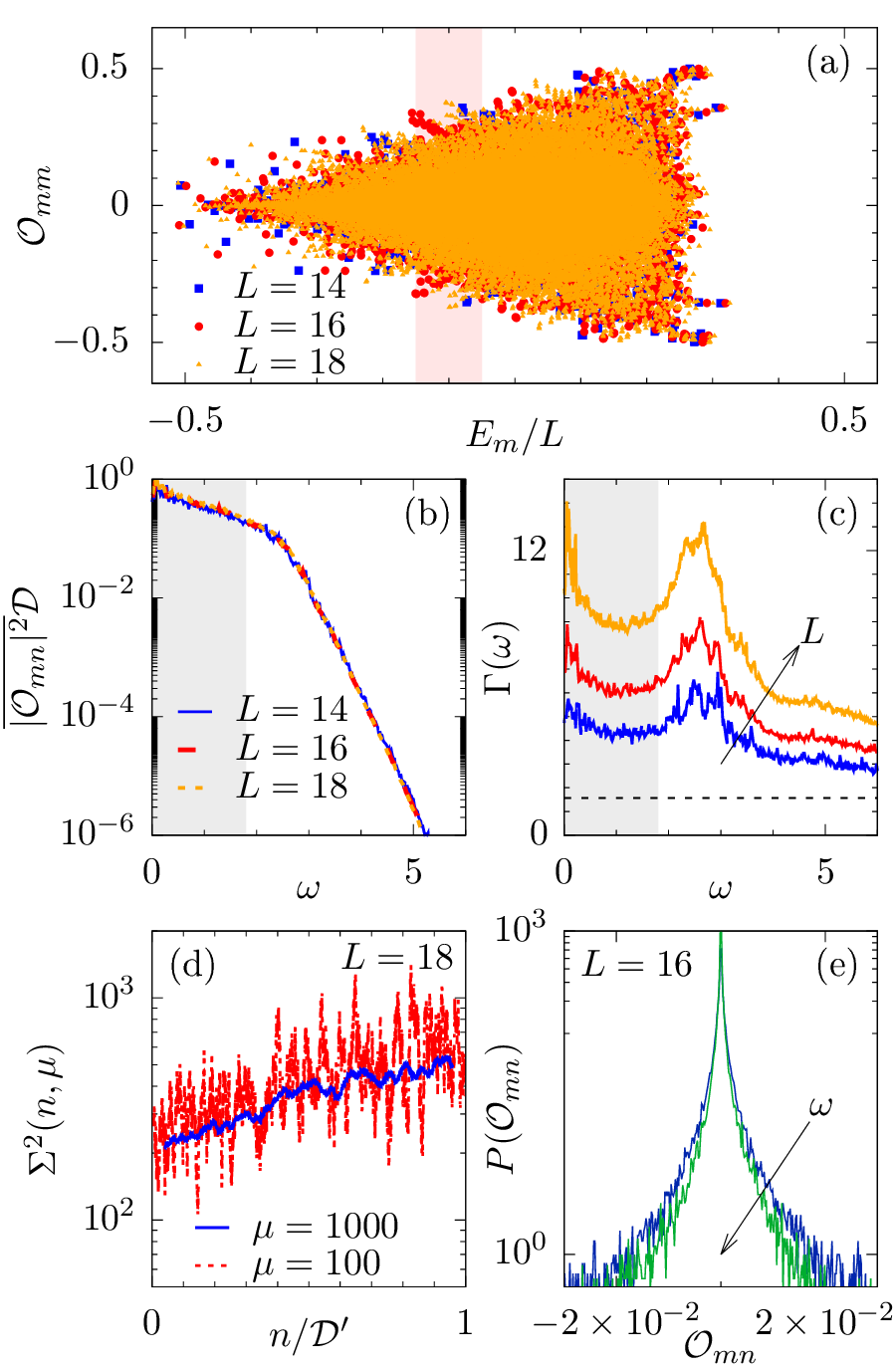}
 \caption{(Color online) Data for the integrable model ${\cal H}_\text{XXZ}$, 
analogous to the results for ${\cal H}_{1,2}$ in Figs.\ \ref{Fig3} - 
\ref{Fig6} of the main text. (a) Diagonal matrix elements ${\cal O}_{mm}$ 
versus $E_m/L$. [(b),(c)] $\overline{|{\cal 
O}_{mn}|^2} {\cal D}$ and $\Gamma(\omega)$ for $L = 14,16,18$. (d) 
$\Sigma^2(n,\mu)$ for $L = 18$ and $\mu = 100,1000$. (e) Distribution 
$P(\mathcal{O}_{mn})$ of off-diagonal matrix elements with $\omega = 
0.1,2$ (arrow). Data in panels (b) to (e) are obtained in the energy 
window $\bar{E}/L \in [-0.15,-0.05]$.}
 \label{Fig11}
\end{figure}
In Sec.\ \ref{Sec::EVs}, we have demonstrated that the eigenvalue 
distribution of band 
submatrices approximately takes on a semicircle shape when the band is 
sufficiently narrow. In Fig.\ \ref{Fig10} 
we show a qualitatively similar result for the Hamiltonian ${\cal H}_1$ 
and full (nonband)  
submatrices, where the width of the energy window is now chosen as $\Delta E/L 
\approx 0.06$ and $\Delta E/L \approx 0.012$, i.e., narrower than in the main 
text. Recall that a smaller $\Delta E$ implies a smaller submatrix 
dimension 
${\cal D}'$.
For the examples shown here, we have 
${\cal D}^\prime \approx 8000$ and ${\cal 
D}^\prime \approx 1500$.
While $P_{\omega_c}(\lambda)$ still 
exhibits pronounced features at $\lambda = \pm 1/2$ for the larger $\Delta 
E$ in Fig.\ \ref{Fig10}~(a), we find that the distributions 
$P_{\omega_c}(\lambda)$ and 
$\widetilde{P}_{\omega_c}(\lambda)$ are essentially indistinguishable for the 
smaller $\Delta E$ in Fig.\ \ref{Fig10}~(b). Moreover, 
analogous to the results for the band submatrices in 
Fig.\ \ref{Fig8}, small deviations from a perfect semicircle 
law appear at the spectral edges if $\Delta E$ is lowered.
While we cannot entirely 
exclude the possibility of  
finite-size effect, we conclude that our results for band
submatrices in the main text, i.e., a semicircular bulk with small deviations 
at the spectral edges, are not just caused by the finite bandwidth, but are 
stable features which appear for full nonband matrices as well.

\section{Results for the integrable model ${\cal H}_\text{XXZ}$}\label{App::Int}

\begin{figure}[tb]
 \centering
 \includegraphics[width=\columnwidth]{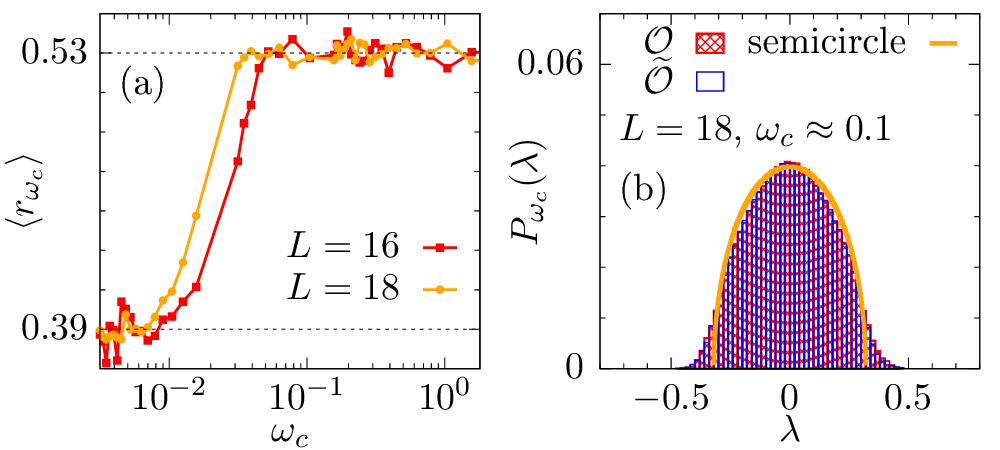}
 \caption{(Color online) Properties of eigenvalues of band submatrices for the 
integrable model ${\cal H}_\text{XXZ}$. Analogous to the results presented in 
the main text, the energy window is chosen as $\bar{E}/L \in [-0.15,-0.05]$. 
(a) $\langle r_{\omega_c} \rangle$ for 
$L = 16,18$. (b) $P_{\omega_c}(\lambda)$ and 
$\widetilde{P}_{\omega_c}(\lambda)$ for $L = 18$ and $\omega_c \approx 0.1$.}
 \label{Fig12}
\end{figure}

In Figs.\ \ref{Fig3} - \ref{Fig6} of the main text, we have analyzed the ETH 
structure of $S_{L/2}^z$ written in the eigenstates of the two nonintegrable 
models ${\cal H}_1$ and 
${\cal H}_2$. In Fig.\ \ref{Fig11},  we present analogous data for the 
integrable model ${\cal H}_\text{XXZ}$. As expected, the results for ${\cal 
H}_\text{XXZ}$ are 
drastically different compared with  the nonintegrable systems  ${\cal 
H}_{1,2}$. In 
particular: (i) the width of the distribution of the diagonal matrix elements 
does not visibly shrink with increasing $L$, (ii) $\Gamma(\omega) \neq \pi/2$ 
and is nonconstant and dependent on $L$, in agreement with \cite{LeBlond2019}, 
(iii) the ratio $\Sigma^2(n,\mu)$ is orders 
of 
magnitude larger compared to $\Sigma^2_\text{GOE} = 2$, and (iv) 
the distribution $P({\cal O}_{mn})$ is clearly non-Gaussian (see also 
\cite{beugeling2015, LeBlond2019}). Overall, these 
results confirm the expectation that the ETH is not satisfied in the case of 
integrable models.  
The only quantity which exhibits a similar behavior for ${\cal H}_\text{XXZ}$ 
and ${\cal H}_{1,2}$ is the running average $\overline{|{\cal 
O}_{mn}|^2}$ shown in Fig.\ \ref{Fig11}~(b). Namely, we find that plotting 
$\overline{|{\cal 
O}_{mn}|^2}$ versus $\omega$ for system sizes $L = 14,16,18$ yields smooth 
curves which collapse onto each other when rescaled by the respective 
Hilbert-space dimension ${\cal D}$. This  behavior is in agreement with 
the recent studies in Refs.\ \cite{LeBlond2019, Mallayya2019}.

Finally, we consider eigenvalue distribution for the
band submatrices in the case of
integrable model ${\cal H}_\text{XXZ}$. Figure
\ref{Fig12}~(a) shows results for the level-spacing ratio $\langle 
r_{\omega_c} \rangle$, while Fig.\ \ref{Fig12}~(b) shows the eigenvalue 
distributions $P_{\omega_c}(\lambda)$ and $\widetilde{P}_{\omega_c}(\lambda)$ 
of the original and the randomized band submatrices with the cutoff frequency 
$\omega_c 
\approx 0.1$. Comparing with the results for the nonintegrable models ${\cal H}_{1,2}$, 
the qualitative behavior of both $\langle r_{\omega_c} \rangle$ and 
$P_{\omega_c}(\lambda)$ appears to be very similar. 
Namely, we find that $\langle r_{\omega_c} \rangle$ exhibits a crossover from 
$r_\text{GOE}$ to $r_\text{Poisson}$ when the cutoff frequency 
$\omega_c$ (or bandwidth $W$) decreases.
Furthermore, $P_{\omega_c}(\lambda)$ and 
$\widetilde{P}_{\omega_c}(\lambda)$ have the very similar shape for the considered value of 
$\omega_c$, which indicates that ${\cal 
O}_{mn}$ within the corresponding band can be considered as independent.  
Accordingly, as discussed in the Appendix \ref{App::RMT}, eigenvalue 
distribution is approximately  semicircular.
Given these results, we conclude that mutual independence of the off-diagonal 
matrix elements below certain energy scale $\omega < \Delta 
E_{\rm 
RMT}$ is also present in the integrable models, raising the question if an 
appropriate non-Gaussian random matrix theory can capture the universal 
properties of the ${\cal O}_{mn}$ in this case.

\section{Eigenvalue distribution of band random matrices}\label{App::RMT}
In this section we briefly review the work of Molchanov, Pastur, Khorunzhii 
\cite{Molchanov}, which derives an integral equation  satisfied by the 
eigenvalue distribution function of a band random matrix. Namely we consider an 
${\cal D}\times {\cal D}$ matrix 
\begin{equation}
{\cal O}_{nm}={v(t)\over \sqrt{{\cal D}}}\, r_{nm},\qquad t=(n-m)/{\cal D}, 
\label{MPK}
\end{equation}
where $v$ is a piece-wise continuous function and $r_{nm}$ are independently 
distributed random variables with zero mean and unit variance. Notice 
that $r_{nm}$ do not have to be Gaussian, it is sufficient that all $r_{nm}$ 
are drawn from the same distribution. So far band submatrices of local 
operators considered in this paper \eqref{Eq::BandMat} are small enough such 
that the density of states  is approximately constant, they can be modeled by 
the random matrix \eqref{MPK} with 
\begin{eqnarray}
v^2(\omega/\Delta E)=\left\{\begin{array}{cc}(\Delta E) f^2(\bar E,\omega), & 
|\omega|\leq \omega_c,\\
0, & |\omega|>\omega_c.
\end{array}\right.
\end{eqnarray}
where $\Delta E=E_{\cal D}-E_1$.
The resolvent 
\begin{equation}
r(z)={\rm Tr}{1\over z-{\cal O}} 
\end{equation}
can be expressed in terms of an auxiliary function $r(t,z)$, $|t|\leq 1/2$ which 
satisfies 
\begin{eqnarray}
&&r(z)=\int_{-1/2}^{1/2} r(t,z) dt,\\
&&r(t,z)\left(z+\int_{-1/2}^{1/2}v^2(t-t')r(t',z)dt'\right)+1=0.
\end{eqnarray}
There are two limiting cases which can be solved analytically, square random 
matrix with $v^2={\rm const}$ (this is the case of $\omega_c=\Delta E$) and 
infinitely narrow band matrix $v^2(t)=v_0^2\, \delta (t)$ (this  is the case of 
$\omega_c\ll \Delta E$). In both cases $r(t,z)=r(z)$ is $t$-independent and 
satisfies 
$v_0^2 r^2+z\, r+1=0$ where
\begin{eqnarray}
v_0^2=\int_{-1/2}^{1/2} v^2(t)dt=2\int_0^{\bar \omega} f^2(\bar E,\omega) 
d\omega,\\
\bar{\omega}=\min(\omega_c,\Delta E/2).
\end{eqnarray}
The eigenvalue distribution is then the semicircle of radius $2v_0$,
\begin{eqnarray}
P(\lambda)= \lim_{\epsilon\rightarrow 0^+}{1\over 
\pi}\Im[r(\lambda+i\epsilon)]={\sqrt{4v_0^2-\lambda^2}\over 2\pi v_0^2}.
\end{eqnarray}



\begin{thebibliography}{99}
\bibitem{Note}
The variable $n$ indicates the 
midpoint of the interval, i.e., the sums run
from $n-\mu/2$ to $n+\mu/2$. In the case 
of $\sigma_\text{od}^2(n,\mu)$, the sum over diagonal matrix elements has to 
be 
replaced by a double sum over off-diagonal matrix elements instead.
Since $\Sigma^2(n,\mu)$ can be dependent on the specific 
values of $n$ and $\mu$, we calculate $\Sigma^2(n,\mu)$ for different choices 
of $\mu$ and all possible intervals $n\in [1+\mu/2,{\cal D}'-\mu/2]$ which can 
be embedded along the diagonal of the submatrix.
\end{thebibliography}
\end{document}